\documentclass[epjST]{svjour}
\usepackage{graphicx,amsfonts,amsmath,amssymb,xcolor,subfigure}
\usepackage[english]{babel}
\newcommand{\pt}[2]{\frac{\mathrm{d} #1}{\mathrm{d} #2}\,}
\newcommand{\e}{\varepsilon}

\newcommand{\Z}{\mathbb{Z}}

\newcommand{\V}[1]{\mathbf{#1}}

\newcommand{\intl}[3]{\int\limits_{#1}^{#2}\mathrm{d}#3\,}
\newcommand{\bracket}[1]{\left(#1\right)}

\newcommand{\grad}{\mbox{\boldmath$\nabla$}}

\newcommand{\grb}[1]{\mbox{\boldmath $#1$}}

\newcommand{\eq}[1]{$\mathrm{Eq.}$~\eqref{#1}}
\newcommand{\eqs}[1]{$\mathrm{Eqs.}$~\eqref{#1}}
\newcommand{\figref}[1]{Fig.~\ref{#1}}

\newcommand{\secref}[1]{$\mathrm{Sect.}$~\ref{#1}}

\newcommand{\av}[1]{\left\langle\, #1 \,\right\rangle}

\usepackage{ulem}

\begin{document}
\title{How entropy and hydrodynamics cooperate in rectifying particle transport}
\author{S. Martens\inst{1}\fnmsep\thanks{\email{steffen.martens@physik.hu-berlin.de}}
\and G. Schmid\inst{2}
\and A. V. Straube\inst{1}
\and L. Schimansky-Geier\inst{1}
\and P. H\"anggi\inst{2}}
\institute{Department of Physics, Humboldt-Universit\"{a}t zu Berlin,
Newtonstr. 15, 12489 Berlin, Germany
\and Department of Physics, Universit\"{a}t Augsburg,
Universit\"{a}tsstr. 1, 86135 Augsburg, Germany}

\abstract{
Using the analytical Fick-Jacobs approximation formalism and extensive Brownian dynamics simulations we study particle transport through
two-dimensional periodic channels with triangularly shaped walls. Directed motion is caused by the
interplay of constant bias acting along the channel axis and a pressure-driven flow. In particular, we analyze
the particle mobility and the effective diffusion coefficient. The mechanisms of entropic rectification is revealed  
in channels with a broken spatial reflection symmetry in presence of
hydrodynamically enforced entropic trapping. Due to the combined action of the forcing and the pressure-driven flow
field, efficient rectification with a drastically reduced diffusivity is achieved.
}

\maketitle
\section{Introduction}
\label{sec:introduction}

Particle separation techniques of micro- or even nanosized particles are based on
the fact that the particles' response to external stimuli, such as gradients or fields, depends on their size. Accordingly,
conventional methods for filtering particles involve centrifugal fractionation \cite{Harrison2003}, phoretic forces \cite{Dorfman2010}
or external fields \cite{Macdonald2003}. Recently, novel devices for particle separation were proposed
\cite{HanggiRMP2009,Reguera2012,Bogunovic2012,Kettner2000}. These utilize the ratchet effect, i.e., the directed transport
under non-equilibrium conditions in periodic systems with broken spatial symmetry \cite{ReimannAPA2002,Astumian2002}.
The use of microfluidic channel systems is promising with respect to separation efficiency, speed
and purity \cite{Reguera2012,Kettner2000}.
The transport in channels with periodically
varying walls exhibits peculiar transport phenomena \cite{Reguera2006,Burada2009_CPC,Martens2013} which can be treated
by means of the so-termed Fick-Jacobs formalism and its generalizations
\cite{Jacobs,Zwanzig1992,Reguera2001,Kalinay2006,Kalinay2011,Sokolov2010,Martens2011,Martens2011b}.
The capability of such devices for separation of particles is rooted in the effect of {\itshape entropic rectification}, i.e.,
the rectification of motion caused by broken spatial symmetry \cite{Schmid2009,Rubi2010,Zitserman2011,Dagdug2012}.

Besides the direct forcing of the particle dynamics,
the application of hydrodynamical flows presents an additional ``degree of freedom'' to control particle transport and optimize rectification.
As pointed in a recent preliminary report by us \cite{Martens2013},
upon  combining a constant force causing the particle to move along the channel and a pressure-driven flow that
drags the particle in the opposite direction results the phenomenon of {\itshape hydrodynamically enforced entropic trapping} was described;
it implies that for certain values of the constant force and the pressure drop, the particle mean flux vanishes and, in addition, the particles' diffusivity drastically reduces.
With this work, we address a different problem, namely the topic of an entropic rectification in the presence of a hydrodynamic flow field.

After presenting the model in \secref{sec:modelling}, we apply the generalized Fick-Jacobs theory to the two-dimensional geometry with triangularly shaped confining walls in \secref{sec:fickjacobs}.
We discuss the standard {\itshape entropic rectification} phenomenon in \secref{subsect:ER}. \secref{sec:interplay}
is devoted to the combined action of constant forcing and the hydrodynamical flow field on the rectification phenomenon.
In \secref{sec:conclusion} we  summarize our main findings.


\section{Statement of the problem}
\label{sec:modelling}

\begin{figure}
  \centering
  \includegraphics[width=0.7\textwidth]{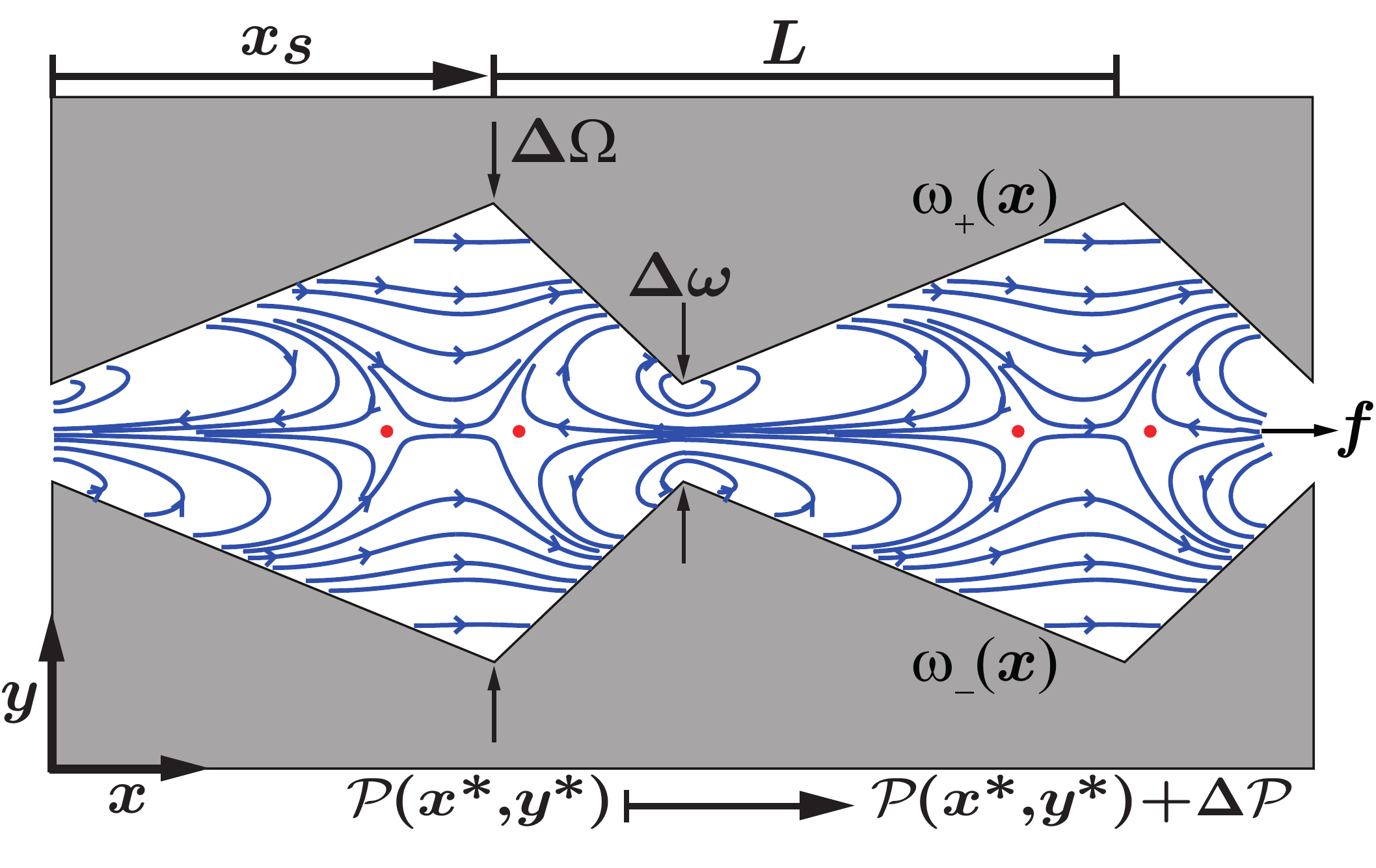}
  \caption{A segment of a triangular shaped channel restricting the overdamped motion of point-like Brownian particles. The zigzag profiles with period $L$
  are given by periodic functions $\omega_{\pm}(x)$, see \eq{eq:linconf}, whose spatial asymmetry is determined by $x_s$.
  The quantities $\Delta \omega$ and $\Delta \Omega$ denote the minimal and maximal channel widths, respectively.
  The (blue) field lines depict a typical external force field $\V{F}(\V{q})$ exerted on the particles, which are
  the result of the interplay of a constant bias $f$ and an oppositely oriented divergence-free force from the driven solvent
  induced by a pressure $\mathcal{P}(\V{q})$ with pressure change $\Delta \mathcal{P}$ along a unit cell.
  This interplay results both in vortices and stagnation points (red circles).
}\label{fig:Fig1}
\end{figure}

We consider spherical Brownian particles of radius $R$ suspended in a solvent with dynamical viscosity $\eta$
in a channel with symmetric confining walls of triangular shape, cf. \figref{fig:Fig1}.
As argued in an earlier account \cite{Martens2013}, we are interested in a two-dimensional (2D) geometry in
which the planar channel is $L$-periodic, with the maximum and minimum widths $\Delta \Omega$ and $\Delta \omega$, respectively,
and the position $x_s$ of the largest width. The confining zigzag walls are described by piecewise-linear boundary functions
\begin{align}\label{eq:linconf}
\omega_\pm(x) = \pm \frac{\e L}{2}
\begin{cases}
    \frac{x}{x_s}+\frac{\delta}{1-\delta}, & \text{for } x \leq x_s \,\, (\text{mod }L)\,,  \\
    \frac{x-1}{x_s-1}+\frac{\delta}{1-\delta}, & \text{elsewhere}\,.
\end{cases}
\end{align}
In \eq{eq:linconf}, $\delta$ denotes the ratio of minimum to maximum channel width,
i.e., $\delta:=\Delta \omega\big/ \Delta \Omega$,
and $\e := (\Delta \Omega - \Delta \omega) \big/ L$ is the dimensionless geometric parameter \cite{Martens2013,Martens2011}.
The channel's asymmetry is controlled by $x_s$: for $x_s \to L/2$ the channel exhibits reflection symmetry;
for $x_s \to 0$ or $x_s \to L$ the largest asymmetry is achieved \cite{Dagdug2012}.
Note that for $\delta \to 1$ and $\e \to 0$ the channel exhibits constant width $\Delta \Omega$, $\omega_\pm(x)=\pm \Delta \Omega/2$.

Assuming throughout (i) dilute particle density inside the channel, (ii) negligible particle size ($R\ll \Delta\omega$), and (iii) a strong viscous dynamics, implies that inertial effects, hydrodynamic particle-particle and particle-wall interactions, and effects that can be initiated by
rotation of particles can largely be neglected \cite{Martens2013,Happel1965,Maxey1983}. If so, the particles evolve in a laminar flow and the dynamics is well
described by the overdamped Langevin equation.
By measuring the lengths, energies, forces, and time in the scales of $L$, $k_BT$, $k_BT/L$, and the relaxation time $\tau=\, 6\pi \eta R L^2/(k_B T)$,
respectively, we obtain the dimensionless Langevin equation describing the particle's motion \cite{Martens2013}
\begin{align}
& \pt{\V{q}}{t} = \V{F}(\V{q}) + \sqrt{2}\,\grb{\xi}(t)\,, \qquad \V{F}(\V{q})=\,-\grad \Phi(\V{q})+\grad\times\grb{\Psi}(\V{q})\,,
\label{eq:eom}
\end{align}
where ${\V{q}=(x,y)^T}$ is the particle position, $k_BT$ is the thermal energy, and the Gaussian random force $\grb{\xi}=(\xi_x,\xi_y)^T$ obeys
$\av{\xi_i(t)} = 0$, $\av{\xi_i(t) \xi_j(s)}=\delta_{ij}\delta(t-s)$; $i,j \in \left\{x,y\right\}$.
In general, any force field $\V{F}$
exerted on particles can be decomposed into a curl-free part (scalar potential $\Phi$) and a
divergence-free part (vector potential $\grb{\Psi}$), which constitute the two components of
the Helmholtz's decomposition theorem.

Hereafter, we study the impact of both contributions to the force $\V{F}$ in \eq{eq:linconf} on the particle transport
through the described geometry. On the one hand, we consider an external constant bias in $x$-direction with magnitude $f$, leading to $\Phi(\V{q})=-f\,x$. On the other hand, we account for the difference between the particle velocity
$\dot{\V{q}}$ and the local instantaneous velocity of the solvent $\V{u}(\V{q})$ based on the Stokes law, which gives us
$\V{u}(\V{q}) = \grad\times\grb{\Psi}(\V{q})$ with $\grb{\Psi}(\V{q}) = \Psi(\V{q}) \V{e}_z$.
Here, $\Psi(\V{q})$ is the hydrodynamic stream function. As a result, \eq{eq:eom} turns into the Langevin equation
\begin{align}
  \pt{\V{q}}{t}=\,f\V{e}_x+\V{u}(\V{q})+\sqrt{2}\,\grb {\xi}(t)\,,
\label{eq:eom_flow}
\end{align}
to be supplemented by no-flux boundary conditions for the particles at the walls.

\section{Generalized Fick-Jacobs approach}
\label{sec:fickjacobs}

%
We next present our generalized Fick-Jacobs approach \cite{Martens2013} which extends the standard Fick-Jacobs theory
towards the most general force fields $\V{F}$ as detailed with \eq{eq:eom}. We first compute the joint probability density
function (PDF) $P\bracket{\V{q},t}$ of finding the particle at the local position $\V{q}$ at time $t$, provided
it was at position $\V{q}=\bracket{0,0}^T$ at time $t=0$. The evolution of $P\bracket{\V{q},t}$ is governed by the
Smoluchowski equation \cite{Burada2009_CPC,50yKramers}
\begin{align}
 \partial_t P(\V{q},t)=\,-\grad\left[\V{F}(\V{q})P(\V{q},t)\right]+\grad^2 P(\V{q},t)\,, \label{eq:smol}
\end{align}
supplemented by the no-flux boundary conditions at the channel's walls. In the long time limit, $\lim_{t\to \infty}
P\bracket{\V{q},t}=P\bracket{\V{q}}$, the PDF has to satisfy the normalization condition $\int_{\mathrm{unit-cell}}^{}
P(\V{q})\, \mathrm{d}^2\V{q} =1$, and be periodic, $P(x+m,y)=P(x,y)\,,\forall m \in \Z$.

Assuming fast equilibration in the transverse channel direction, we perform an asymptotic perturbation analysis in the geometric parameter
$\e \ll 1$ \cite{Martens2011,Martens2011b}, cf. Refs. \cite{Kalinay2006,Kalinay2011,Laachi2007}.
Upon re-scaling the transverse coordinate $y\to \e\,y$, the profile functions and the vector potential
become $\omega_\pm(x)\to \e\,h_\pm(x)$ and $\Psi\to \e\,\Psi$, respectively. Expanding the joint PDF
in a series in even powers of $\e$, we get $P(\V{q},t)=P_0(\V{q},t)+\e^2\,P_1(\V{q},t)+O(\e^4)$ and
similarly $\Phi(\V{q})=\Phi_0(\V{q})+O(\e^2)$ and $\Psi(\V{q})=\Psi_0(\V{q})+O(\e^2)$. Substituting
this ansatz into \eq{eq:smol} and observing the boundary conditions, we obtain a hierarchic set of
partial differential equations. For the steady state, we find $P_0(\V{q})=g(x)\,\exp(-\Phi_0(\V{q}))$,
where $g(x)$ is obtained in the order $O(\e^2)$. With the condition that the $x$-component of ${\grad\times\grb{\Psi}_0}$
is periodic in $x$ with unit period the stationary marginal PDF, $P_0(x)=\lim_{t\to\infty}\int_{h_{-}(x)}^{h_{+}(x)}\mathrm{d}y\, P_0(\V{q},t)$, yields
\begin{align}
\label{eq:pdf}
P_0(x)=\mathcal{I}^{-1}I(x)\,.
\end{align}
Here, $I(x) = e^{-{\cal F}(x)} \int_{x}^{x+1} \mathrm{d}x' e^{{\cal F}(x')}$, with $\mathcal{I} = \int_{0}^{1} \mathrm{d}x
\,I(x)$, and
$\mathcal{F}(x)$ is the generalized potential of mean force, reading
\begin{align}
 \label{eq:meanforce}
\begin{split}
  \mathcal{F}(x)&=\,-\ln\left[\intl{h_-(x)}{h_+(x)}{y}\,
e^{-\Phi_0(\V{q})}\right] -\intl{0}{x}{x'}\intl{h_-(x')}{h_+(x')}{y}\,
\bracket{\grad \times \grb{\Psi}_0}_x\,P_\mathrm{eq}(y|x')\,,
\end{split}
\end{align}
with $P_\mathrm{eq}(y|x)=\,e^{-\Phi_0(\V{q})}\Big/\int_{h_-(x)}^{h_+(x)}\mathrm{d}y\, e^{-\Phi_0(\V{q})}$.
We reveal that $\mathcal{F}(x)$ comprises the usual \textit{entropic} contribution (the logarithmic term)
\cite{Reguera2006,Burada2008} caused by the non-holonomic constraint stemming from the boundaries
\cite{Sokolov2010,Martens2012b} and an additional, energetic contribution stemming from
$\grb{\Psi}_0$ \cite{Martens2013}.
In the absence of the flow field, i.e., for $\grb{\Psi}=0$, \eqs{eq:pdf} and
\eqref{eq:meanforce} reduce to the commonly known result of the \textit{Fick-Jacobs} approximation
\cite{Burada2009_CPC,Zwanzig1992}.

The kinetic equation for the time-dependent marginal PDF $P_0(x,t)$, with the steady-state solution in \eq{eq:pdf}, is
the generalized Fick-Jacobs equation
\begin{align}
  \frac{\partial}{\partial t} P_0(x,t) = \frac{\partial}{\partial x}
\left[ \frac{\mathrm{d} \mathcal{F}(x)}{\mathrm{d} x}\,P_0(x,t)\right]+\frac{\partial^2}{\partial x^2} P_0(x,t)\,. \label{eq:kinetic-eq}
\end{align}
Based on \eq{eq:kinetic-eq}, the stationary mean particle current for our problem
reads \cite{Martens2013}
\begin{align}
  \label{eq:generalized_stratonivich}
 \av{\dot{x}}& = \lim_{t \to \infty} \frac{x(t)}{t} = \,\mathcal{I}^{-1}\,\bracket{1-e^{\Delta\mathcal{F}}},
\end{align}
wherein  $\Delta\mathcal{F}=\mathcal{F}(x+1)-\mathcal{F}(x)$. Note that \eq{eq:generalized_stratonivich}
is a generalization of the well-known Stratonovich formula \cite{Stratonovich} which
was originally derived for tilted periodic potentials.

\section{Entropic rectification}
\label{subsect:ER}

Here, we shall focus on transport across  triangular-shaped channel structures induced by a curl-free force field, $\V{F}(\V{q},t)=f$, in a resting solvent, $\V{u}(\V{q},t)=0$. Accordingly, the generalized potential of mean force \eq{eq:meanforce} simplifies to the  ``entropic'' potential $\mathcal{F}(x)= A(x)=-f\,x-\ln\left[2\omega(x)\right]$ \cite{Burada2009_CPC,Burada2008}.

\begin{figure}
 \subfigure{\includegraphics[width=0.5\textwidth]{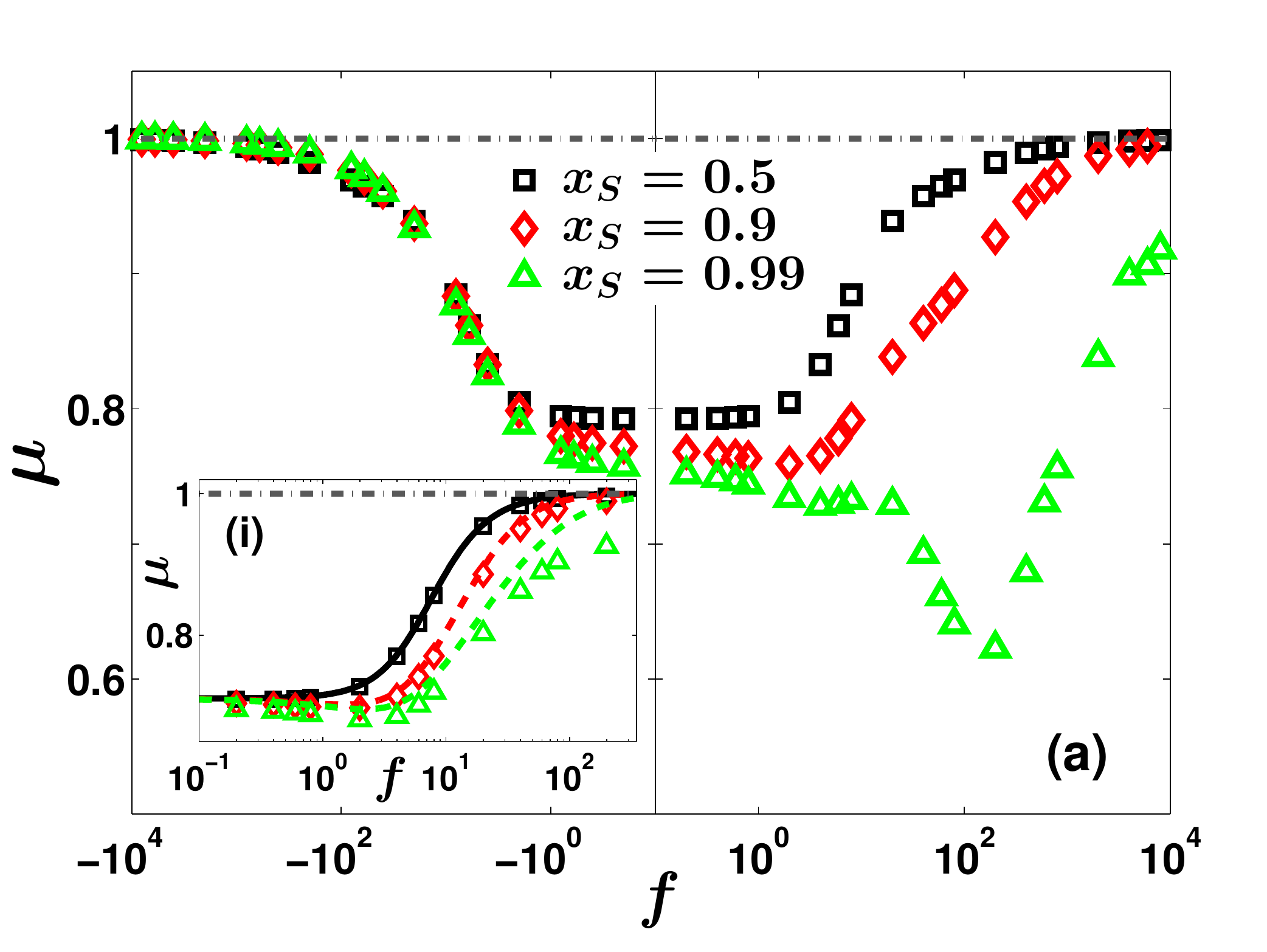}}
 \subfigure{\includegraphics[width=0.5\textwidth]{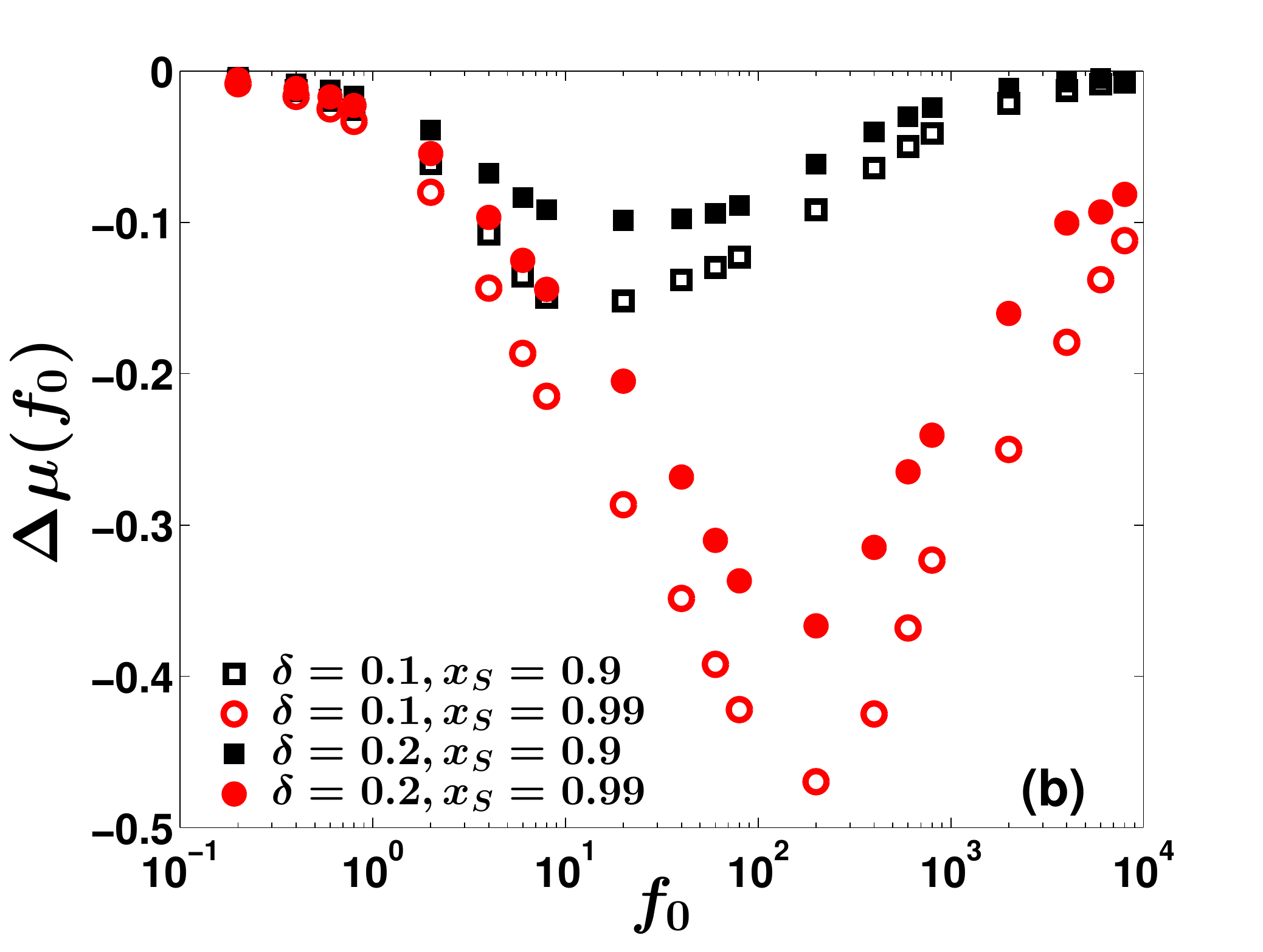}}
 \caption{\label{fig:Fig2} Numerical results (markers) for the
 particle mobility $\mu(f)$ (a) and mobility difference $\Delta \mu(f_0)=\mu(f_0)-\mu(-f_0)$ (b)
 versus the force magnitude $f$ and $f_0$, respectively, for a quiescent solvent, $\V{u}=0$.
 For panel (a), the asymmetry parameter $x_s$ is varied, while the channel widths are fixed,
 $\Delta \omega=0.1$ and $\Delta \Omega=0.5$. Note the logarithmic scale for the whole range of $f$.
 The horizontal dash-dotted line indicates unity. Inset (i): Comparison of Brownian dynamics simulations
 (markers) of \eq{eq:eom_flow} with the Fick-Jacobs approximation (lines), \eq{eq:moblin_FJ}, for a
 channel with $\Delta\Omega=0.1$ and $\Delta\omega=0.01$. In panel (b), the aspect ratio $\delta$
 and the asymmetry parameter $x_s$ are varied, while $\Delta\Omega=0.5$. }
\end{figure}

We first derive the dependence of the average particle velocity $\av{\dot{x}}$ on $f$ and channel's parameters. According to \eq{eq:generalized_stratonivich}, we obtain
\begin{align}
 \label{eq:moblin_FJ}
\av{\dot{x}}^{-1}&=\frac{1}{f}+\frac{\bracket{1-e^{f(x_s-1)}}}{ (1-x_s) f^2 \bracket{1-e^{-f }}} e^{\frac{x_s\delta f}{1-\delta}}\,\Big[\Gamma\bracket{0,\frac{x_s \delta f}{1-\delta}}-\Gamma\bracket{0,\frac{x_s f}{1-\delta}}\Big] \notag
\\ &+\frac{\bracket{1-e^{f x_s}}}{x_s f^2 \bracket{1-e^{-f}}} e^{\frac{(x_s \delta-1) f}{1-\delta}}\Big[\Gamma\bracket{0,\frac{(x_s-1) \delta f}{1-\delta}}-\Gamma\bracket{0,\frac{(x_s-1) f}{1-\delta}}\Big]\,,
\end{align}
for the wall profiles \eq{eq:linconf}, where $\Gamma\bracket{n,a}=\int_{a}^{\infty}\mathrm{d}t\,e^{-t} t^{n-1}$ is the upper incomplete gamma function.
Another important transport quantity is the particle mobility $\mu$ reading
\begin{align}
 \label{eq:mobility}
 \mu := \frac{\av{\dot{x}}}{f},
\end{align}
for any non-zero force $f$. In the case of infinitely large force strength $f$, the particle velocity converges
to its free value, $\lim_{\pm f\to \infty}\av{\dot{x}}=\pm f$, and therefore the particle mobility $\mu$ tends to unity, regardless of the value of the asymmetry parameter $x_s\in (0,1)$.
However, for a wedge-like shaped unit-cell, i.e., $x_s=0$ or $x_s=1$, the particle mobility monotonically decreases with increasing $|f|$ and
converges to the asymptotic value $\mu(f \to -\infty) = \delta$ for $x_s=0$ and $\mu(f \to \infty) = \delta$ for $x_s=1$ \cite{Marchesoni2009} (not explicitly shown in \figref{fig:Fig2}).
Referring to the Sutherland-Einstein relation, the effective diffusion coefficient $D_\mathrm{eff}(f):=\lim_{t \to \infty} (\av{x^2}-\av{x}^2)\big/2t$ coincides
with the particle mobility in the diffusion dominated regime, i.e., $f\ll 1$. For $f\to 0$ of \eq{eq:moblin_FJ}, we derive
\begin{align}
 \lim_{f\to 0} \mu_0(f)=\lim_{f\to 0} D_\mathrm{eff}\bracket{f}=\frac{2\,\bracket{1-\delta}}{\bracket{1+\delta}\,\ln\bracket{1/\delta}}. \label{eq:moblin_FJ_f0}
\end{align}
Interestingly, within the presented Fick-Jacobs theory, the asymptotic value is only determined by the channel's aspect ratio $\delta=\Delta\omega/\Delta\Omega$ and,
more importantly, it is independent of the asymmetry parameter $x_s$.

In \figref{fig:Fig2} (a), we depict the dependences of the particle mobility $\mu$ on the magnitude of the bias $f$
and its direction (positive and negative). The numerical results are obtained by Brownian dynamics simulation of \eq{eq:eom}
for $3\cdot 10^4$ individual particles. Note that there is a strong dependence on the direction of forcing \cite{Schmid2009,Dagdug2012,Kosinska2008,Reguera2010}
reflecting a broken symmetry of the channel (with respect to $x$). For $x_s\neq1/2$, the channel's symmetry is broken, resulting in $\mu(f)\neq\mu(-f)$.
Starting out from a value smaller than the bulk value, i.e., $\mu(f) <1$, the nonlinear mobility monotonically increases with the strength of the bias $|f|$ and approaches unity for $|f|\to \infty$.
However, for symmetric triangular channels corresponding to $x_s=1/2$, the particle mobility does not depend on the direction of the forcing and one observes $\mu(f) = \mu(-f)$.

If the force points into the direction of the gently rising flanks, i.e., for $f<0$ and $x_s>0.5$, the
focusing character of the channel rules the propagation of the particle in channel direction while diffusion in transversal channel direction, which suppresses the mobility in channel direction, is reduced, cf. comparison of time scales in Ref.~\cite{Burada2008}. Consequently, the nonlinear mobility monotonically increases with $f$ and approaches the free value for $f \to -\infty$, cf. \figref{fig:Fig2} (a).

Interestingly, the dependence of $\mu$ on the forcing parameter exhibits a peculiar non-monotonic behavior if the forcing acts towards the direction of the steep flanks, $f>0$.
Due to the steepness of the confining walls the force component directing the particles towards the bottleneck is rather small, $f_t\propto (1-x_s)$.
Therefore the particles' residence time in one unit-cell increases, respectively, the particle mobility goes down with growing asymmetry parameter $x_s$. Similar to the case of septate channels, $x_s \to 1$, the particles can diffusively explore the full $y$-range which results in a pronounced minima of the nonlinear mobility for moderate forcing strengths. Nevertheless, as the focusing effect strengths more and more with increasing bias, the mobility tends to unity for $f\to \infty$.

In the inset of \figref{fig:Fig2} (a) we compare the numerical obtained results for $\mu$ (markers) with the Fick-Jacobs approach result (lines), \eq{eq:moblin_FJ},
for a weakly corrugated channel with $\Delta\Omega=0.1$ and $\Delta\omega=0.01$. Noteworthy, our exact analytic result matches
very well the numerics for almost all values of $f$  and $x_s$. Deviations between \eq{eq:moblin_FJ} and the numerics occur only for extreme asymmetric channel
geometries, i.e., for $x_s \to 1$. In contrast to the usually studied sinusoidally modulated channel \cite{Martens2013,Martens2011,Burada2008,Burada2007}, the requirement $\e\ll 1$ is not sufficient to ensure the validity of the Fick-Jacobs approximation (leading order in $\e$).
Concurrently, as an analysis of the different involved time scales predicts, the cross-section variation rate $\lambda=(1-x_s)/\e$ \cite{Dagdug2011}
has to be small as well. Note that in contradiction to \eq{eq:moblin_FJ_f0} (zeroth order perturbation theory), our numerical simulations show
a weak dependence of $\lim_{f\to 0} \mu(f)$ on the asymmetry parameter $x_s$.


We next discuss the impact of the channel's asymmetry, $0.5\leq x_s < 1$, on the entropic rectification properties.
We assume that the external bias instantly switches between the two values $\pm f_0$ after half a period $T$, i.e., $f(t)=\pm f_0$.
Similar to the Brownian motor phenomenon \cite{HanggiRMP2009,ReimannAPA2002,Astumian2002}, such  oscillating forcing results in directed motion
along the $x$-axis. Assuming adiabatic driving, i.e., the period is much longer than any other involved time scale,
the effective drift velocity is given by $\av{\av{\dot{x}}}_T := 1/T \int_0^T dt\, \av{\dot{x}(f(t))}$ and reduces for dichotomic, unbiased driving to
\begin{align}
 \av{\av{\dot{x}(f_0)}}_T=\,\frac{1}{2}\bracket{\av{\dot{x}(f_0)}+\av{\dot{x}(-f_0)}}=\,\frac{1}{2}\Delta\mu(f_0)\,f_0,
\end{align}
where $\Delta \mu(f_0):=\mu(f_0)-\mu(-f_0)$ denotes the mobility difference for forward and backward driving. Note that the relative mobility difference $\alpha=|\Delta \mu(f_0)| / \bracket{\mu(f_0) + \mu(-f_0)}$ for quantifying the rectification  was introduced earlier as another measure for entropic rectification \cite{Schmid2009}.

In \figref{fig:Fig2} (b), we present $\Delta \mu(f_0)$ for different channel asymmetries $x_s$ and aspect ratios $\delta$.
While for symmetric channels, $x_s=0.5$, the mobility difference equals zero regardless the value of $f_0$, $\Delta \mu(f_0)=0$, a non-monotonic behavior is
observed if the channel symmetry is broken. For $0.5 < x_s < 1$, as $\mu(-f_0) \geq \mu(f_0)$ the mobility difference $\Delta \mu$ is negative and thus the particles' motion is rectified, viz. they move with $\av{\av{\dot{x}}}_T<0$ to the left. For the limiting cases $f_0 \to \infty$ and $f_0 \to 0$, the forward and backward mobility equals each other, cf. \figref{fig:Fig2} (a), and correspondingly $\Delta \mu(f_0)$ approaches zero. Note that both the minimum value of $\Delta \mu(f_0)$ decreases and the position of the minimum shifts to larger force magnitudes $f_0$ with growing channel asymmetry. Solely for wedge-like shaped unit-cells, $x_s=1$, $\Delta \mu(f_0)$ monotonically decreases with $f_0$ from zero to the asymptotic value $\Delta\mu(\infty)=(\delta-1)$. Decreasing the bottleneck width, respectively, the aspect ratio $\delta$ results in a systematic shift of all curves towards smaller values for $\Delta \mu(f_0)$. Simultaneously, the value of $f_0$ where the mobility difference is maximal remains.

As different sized particles may be treated by consideration of an effective bottleneck widths \cite{Riefler2010} and given that the mobility difference exhibits a strong dependence on $\delta$, our channel setup is an ideal candidate for a device separating particles according to their sizes \cite{Reguera2012}. Since the particle transport is accompanied by a strong enhancement of the effective diffusivity \cite{Dagdug2011}, the quality of entropic particle rectification may be weak. Recently, we found that the counteraction of a pressure-driven flow and a constant bias of strength $f$ ensues the intriguing {\itshape hydrodynamically enforced entropic trapping} (HEET) phenomenon where the vanishing of the mean particle current is accompanied by a significant suppression of diffusion \cite{Martens2013}. Consequently, the quality of hydrodynamically induced particle rectification may be better.

\section{Interplay of pressure-driven solvent flow and external bias}
\label{sec:interplay}

Next, we reveal the role of a flow field $\V{u}(\V{q})=\grad\times\grb{\Psi}(\V{q})$ on the particle transport through the triangular channel structures, presenting an example of the divergence-free force.
The dynamics of particles is described by \eq{eq:eom_flow}, which implies an one-way coupling between the solvent and the particles; for a dilute suspension, the fluid flow influences the particle dynamics but not \textit{vice versa} \cite{Straube2011}.

A slow viscous steady flow of an incompressible solvent is governed by the dimensionless Stokes or ``creeping flow'' equations \cite{Happel1965,Bruus2008},
\begin{align}
\grad \mathcal{P}(\V{q})=\grad^2 \V{u}(\V{q})\,, \qquad \grad
\cdot \V{u}\bracket{\V{q}}=0\,, \label{eq:creeping}
\end{align}
being valid for small Reynolds number ${\rm Re}=\rho\,L^2/(\eta\,\tau)\ll 1$. The flow velocity ${\V{u}=\bracket{u_x,u_y}^T}$ and the pressure $\mathcal{P}(\V{q})$
are measured in the units of $L/\tau$ and $\eta/\tau$, respectively. Further, we require that $\V{u}$ obeys periodicity, $\V{u}(x,y)=\V{u}(x+1,y)$,
and the no-slip boundary conditions, $\V{u}(\V{q})=0$, $\forall\, \V{q} \in \mbox{channel wall}$.
The pressure satisfies $\mathcal{P}(x+1,y)=\mathcal{P}(x,y)+\Delta \mathcal{P}$, where $\Delta \mathcal{P}$ is the pressure drop along one unit cell.

Applying the curl to both sides of first relation in \eq{eq:creeping} eliminates $\mathcal{P}(\V{q})$, yielding the biharmonic equation $\grad^4 \Psi(x,y)=0$
for the stream function $\Psi$, related to the flow velocities as ${u_x=\partial_y \Psi}$ and
${u_y=-\partial_x \Psi}$. Following the asymptotic procedure described in \secref{sec:fickjacobs},
%
%
we solve the biharmonic equation ${0=\partial_y^4 \Psi_0(x,y)+O(\e^2)}$ with the no-slip boundary conditions, ${\partial_y \Psi_0=0}$ at the channel walls ${y=h_{\pm}(x)}$, and the conditions specifying the flow throughput $Q=-\Delta
\mathcal{P}/(12 \langle \mathcal{H}^{-3}(x)\rangle_x)$ \cite{Kitanidis1997}, ${\Psi_0=0}$ at ${y=h_{-}(x)}$ and $\Psi_0=Q$ at $y=h_{+}(x)$, to obtain in leading order in $\e$:
\begin{align} \label{eq:stream0}
 \Psi_0=&\,-\frac{\Delta
\mathcal{P}}{12}\,\frac{[y-h_-(x)]^2\left[3h_+(x)-h_-(x)-2y\right]}{\mathcal{H}^3(x)\,\av{\mathcal{H}^{-3}(x)}_x}\,,
\end{align}
where $\mathcal{H}(x)=h_+(x)-h_-(x)$ is the re-scaled local width and $\av{\cdot}_x=\int_{0}^{1} \cdot\,\mathrm{d}x$
denotes the average over one period of the channel.


\subsection{Purely flow-driven transport}

In the absence of conservative forces, when $\Phi(\V{q})=0$ ($f=0$), and for no-slip boundary conditions at the walls, the joint probability density function is spatially uniform,
$P(\V{q})=1/\av{\mathcal{H}(x)}_x$, unless hydrodynamic interactions between particles and walls come into play \cite{Happel1965,Schindler2007}.

Then, for the purely flow-driven case, $f=0$, the mean particle current $\av{\dot{x}}_0^\mathrm{flow}=\av{\mathcal{H}(x)}_x^{-1}\int_{0}^{1}dx \int_{h_-(x)}^{h_+(x)} dy \,u_0^x(x,y)$
is explicitly evaluated for the wall profiles given by \eq{eq:linconf} to give
\begin{align}
 &\av{\dot{x}}_0^\mathrm{flow}=-\frac { \Delta
\mathcal{P}}{12\av{\mathcal{H}}_x\,\av{\mathcal{H}^{-3}}_x}=\,-\frac{\Delta\mathcal{P}
\bracket{\Delta\Omega}^2}{12}\,\frac{2\delta^2\ln\bracket{1/\delta}}{1-\delta^2} \quad (f=0)\,.
\label{eq:curr0_f0}
\end{align}
The mean particle current is as for the Poiseuille flow between plane parallel walls,
$\av{\dot{x}}_\mathrm{\delta=1}=-\Delta\mathcal{P}(\Delta\Omega)^2/12$, modified
by the factor $\kappa(\delta):=2\delta^2\ln\bracket{1/\delta}/\bracket{1-\delta^2}$ accounting for the corrugation;
note that $\lim_{\delta \to 1}\kappa(\delta)=1$. Thus, the solvent flows from left to right for ${\Delta\mathcal{P}<0}$ and \textit{vice versa} for ${\Delta\mathcal{P}>0}$.
Remarkably, result \eq{eq:curr0_f0} is independent of the asymmetry parameter $x_s$. Moreover, since
$\av{\dot{x}(\Delta \mathcal{P})}_0^\mathrm{flow}=-\av{\dot{x}(-\Delta \mathcal{P})}_0^\mathrm{flow}$,
no rectification of point-sized objects can be achieved by periodically switching of the flow's direction -- even for asymmetric channels.
This result, however, has no contradiction with the ``drift ratchet'' \cite{Kettner2000}, where the finite size of the particles was crucial for the transport.


\subsection{Interplay of solvent flow and external bias}


\begin{figure}
 \subfigure{\includegraphics[width=0.5\textwidth]{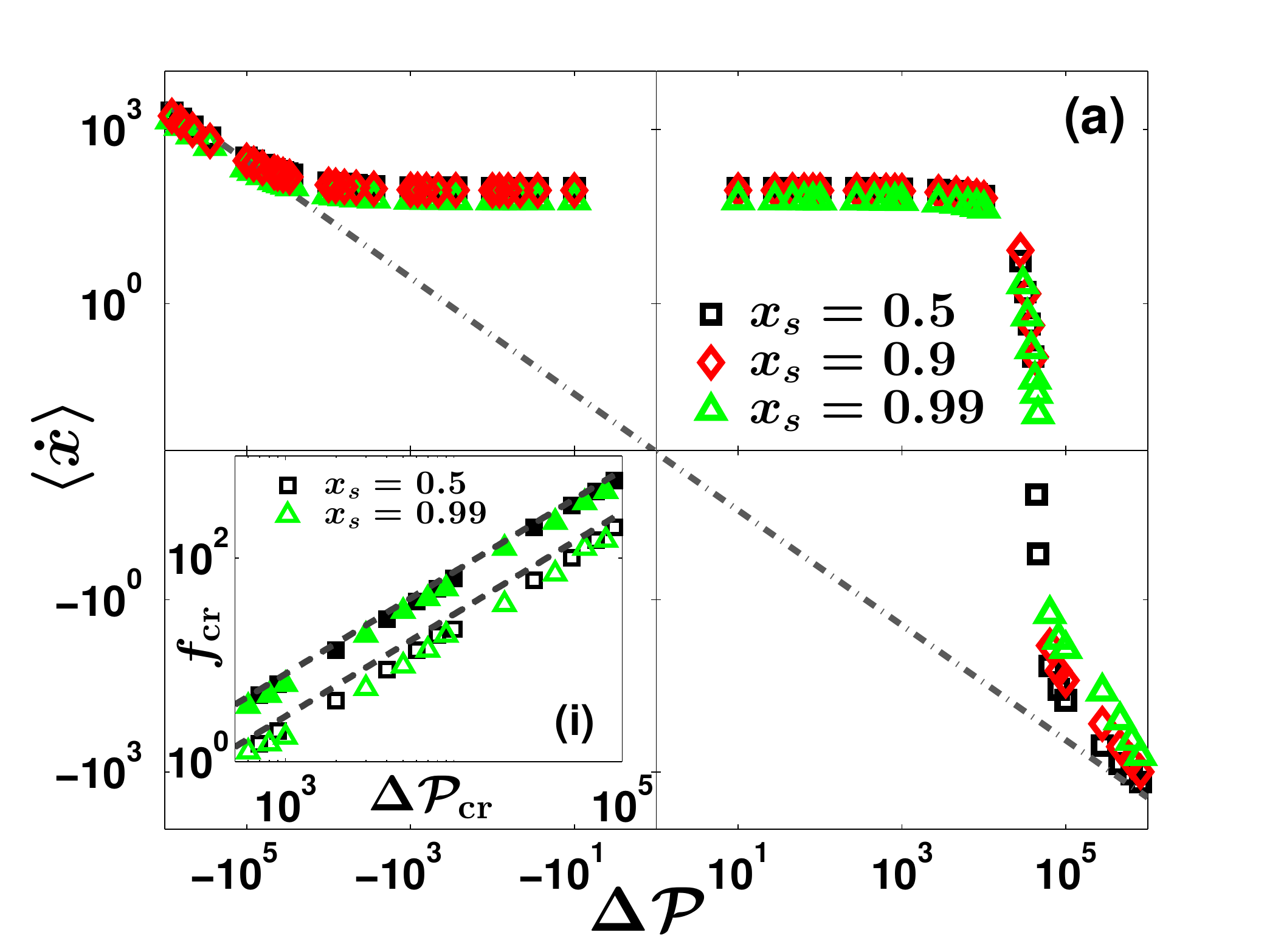}}
 \subfigure{\includegraphics[width=0.5\textwidth]{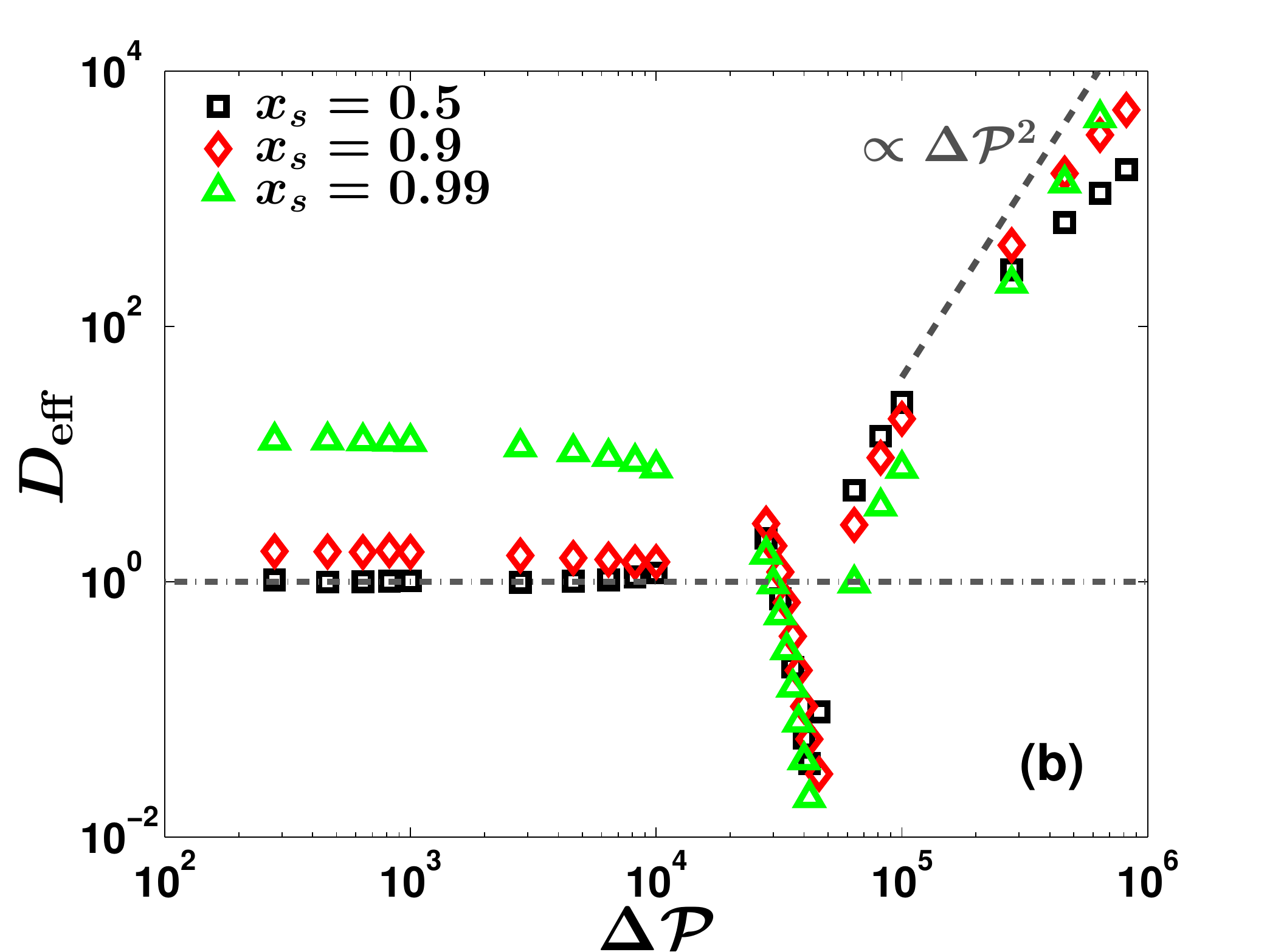}}\\
 \caption{\label{fig:interplay}
 The scaled mean particle velocity $\av{\dot{x}}$ (a) and the effective diffusivity $D_\mathrm{eff}$ (b) as functions
 of the pressure drop $\Delta\mathcal{P}$ for a constant bias $f=100$ and different asymmetry parameters $x_s$ for
 triangularly shaped channels with maximum and minimum widths ${\Delta\Omega=0.5}$ and $\Delta\omega=0.1$, respectively.
 The dash-dotted line in panel (a) indicates the analytic estimate for $f=0$, \eq{eq:curr0_f0}. The horizontal line in
 panel (b) corresponds to the value of the bulk diffusivity. The markers are for Brownian dynamics simulations of \eq{eq:eom} with the flow field obtained via numerically solving \eq{eq:creeping}.
 Inset (i): The dependence of critical force value $f_\mathrm{cr}$ on the critical pressure drop $\Delta \mathcal{P}_\mathrm{cr}$ for different aspect ratios $\delta$ is depicted, viz. $\delta=0.1$ (blank markers) and $\delta=0.2$ (filled markers). The analytic result \eq{eq:f_dp_crit} is shown by the dashed lines.}
\end{figure}

In \figref{fig:interplay}(a), we depict the mean particle velocity $\av{\dot{x}}$ as a function of the pressure drop
$\Delta {\mathcal P}$ for a constant bias $f$ and different values of the asymmetry parameter $x_s$.
While $\av{\dot{x}}$ exhibits the linear dependence predicted by \eq{eq:curr0_f0} for $f=0$,
its behavior changes drastically for $f \neq 0$. For $|\Delta {\mathcal P}| \gg 1$ the drag force dominates over the external bias and thus $\av{\dot{x}} \propto - \Delta {\mathcal P}$. For intermediate values of $\Delta {\mathcal P}$,
$\av{\dot{x}}$ is nearly constant in a broad range of $|\Delta {\mathcal P}|$. For $\Delta {\mathcal P}>0$, the constant bias
drives the particles in the direction opposite to the flow. At a critical pressure drop $\Delta \mathcal{P}_\mathrm{cr}$, a sharp transition
from positive to negative values of $\av{\dot{x}}$ is observed, cf. Ref.~\cite{Martens2013}. As follows from \eq{eq:generalized_stratonivich}, $\av{\dot{x}}=0$ when $F(x+1)-F(x)=\Delta \mathcal{F}=0$, yielding for the critical ratio
\begin{align}
 \bracket{\frac{f}{\Delta
\mathcal{P}}}_\mathrm{cr}=\,\frac{1}{12}\,\frac{\av{\mathcal{H}(x)^{-1}}_x}{\av{\mathcal{H}(x)^{-3}}_x}=\, \frac{(\Delta\omega)^2 \ln(1/\delta)}{6(1-\delta^2)}\,,
\label{eq:f_dp_crit}
\end{align}
being independent of asymmetry parameter $x_s$, cf. \figref{fig:interplay}(i).

The described behavior of $\av{\dot{x}}$ is reflected by peculiar features in the effective diffusion coefficient, cf. \figref{fig:interplay}(b).
If for $|\Delta\mathcal{P}|\ll 1$, $D_\mathrm{eff}$ is mainly determined by the channel's geometry and the constant bias $f$. At
$\Delta\mathcal{P}\approx \Delta\mathcal{P}_\mathrm{cr}$, a drastic reduction of the diffusivity can be observed.
This effect, which occurs when the constant bias and the flow start to counteract such that the field $\V{F}(\V{q})=f \V{e}_x + \V{u}(\V{q})$ contains vortices
and stagnation points and $\av{\dot{x}}\approx 0$, is referred to as the hydrodynamically enforced entropic trapping (HEET).
With the further growth in $\Delta\mathcal{P}$, $D_\mathrm{eff}$ exhibits Taylor-Aris dispersion \cite{Taylor1953,Aris1956} irrespective of the channel constriction, i.e.,
$D_\mathrm{eff}\propto \bracket{\Delta\Omega\,\av{\dot{x}}}^2/ 192$. Thus, the interaction with the flow can significantly affect the macroscopic quantity $D_\mathrm{eff}$.


\begin{figure}[!htb]
 \subfigure{\includegraphics[width=0.5\textwidth]{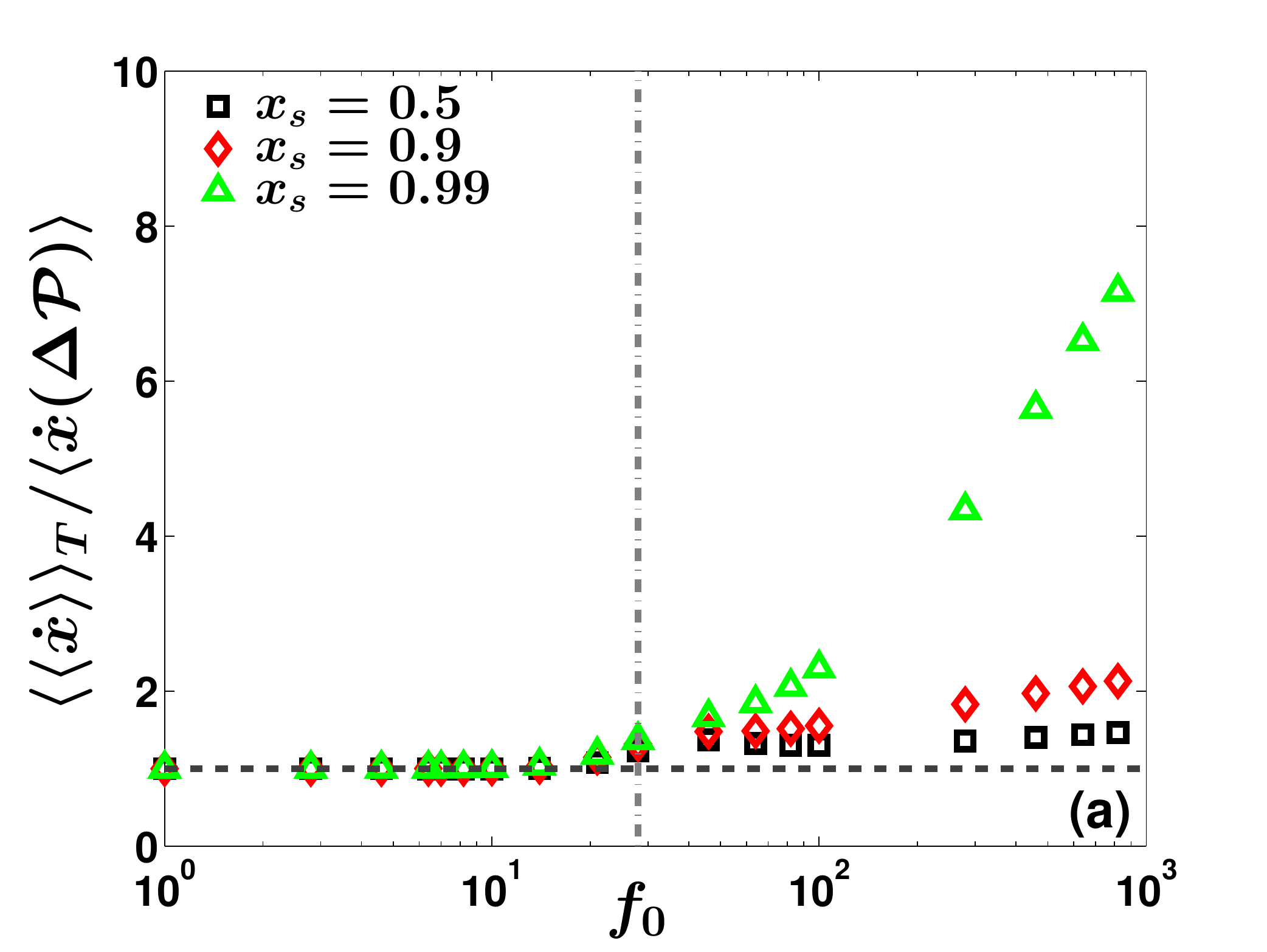}}
 \subfigure{\includegraphics[width=0.5\textwidth]{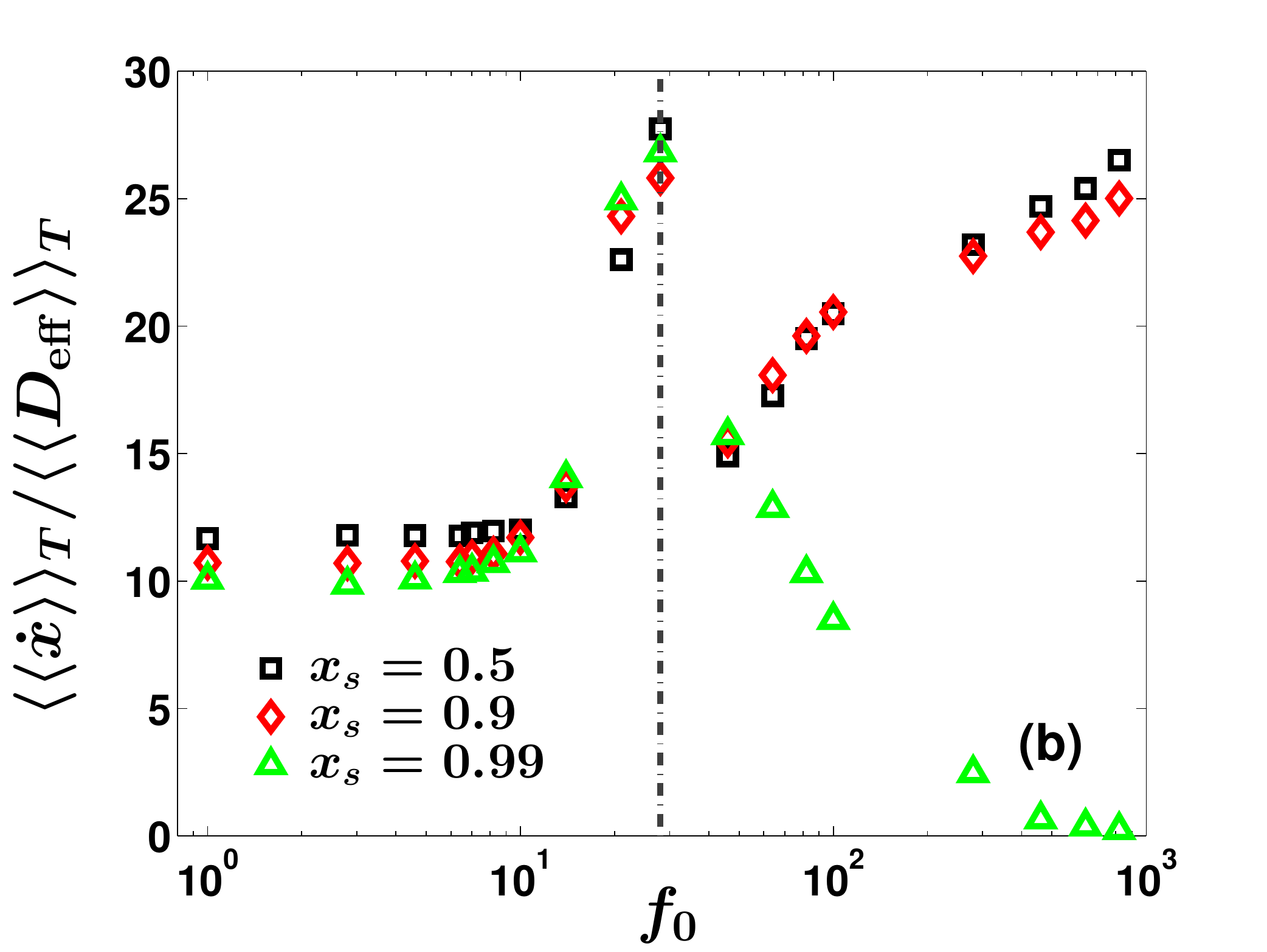}}\\
 \caption{\label{fig:Fig4} Scaled effective drift velocity $\av{\av{\dot{x}}}_T$ (a) and effective P{\'e}clet number $\av{\av{\dot{x}}}_T / \av{D_\mathrm{eff}}_T$ (b) for various compartment shapes. The Brownian particles are driven by a constant pressure drop $\Delta\mathcal{P}=10^4$ and a dichotomic switching bias $f(t)=\pm f_0$. The horizontal dashed line indicates unity, whereas the vertical dash-dotted lines show the value of $f_\mathrm{cr}$, cf. \eq{eq:f_dp_crit}. The channel parameter are set to $\Delta\Omega=0.5$ and $\Delta\omega=0.1$.}
\end{figure}

The strong variation of the mean particle velocity $\av{\dot{x}}$ accompanied by the drastic reduction of diffusivity $D_\mathrm{eff}$
at the critical conditions enables the possibility for efficient rectification. In \figref{fig:Fig4},
we present the impact of the asymmetry parameter $x_s$ on the effective drift velocity $\av{\av{\dot{x}}}_T$
and the effective P{\'e}clet number $\av{\av{\dot{x}}}_T / \av{D_\mathrm{eff}}_T$ for a dichotomic switching bias $f(t)=\pm f_0$.
Since the mean particle velocity is dominated by the Stokes' drag force for $f_0<f_\mathrm{cr}$ the time-averaged
(over period $T$) velocity $\av{\av{\dot{x}}}_T$ coincides with the undisturbed particle velocity $\av{\dot{x}(\Delta {\mathcal P})}$.
As $f_0$ approaches $f_\mathrm{cr}$, $\av{\av{\dot{x}}}_T$ grows and thus $\av{\av{\dot{x}}}_T > \av{\dot{x}(\Delta {\mathcal P})}$
independent of $x_s$. Interestingly, $\av{\av{\dot{x}}}_T > \av{\dot{x}(\Delta {\mathcal P})}$ remains valid for strong bias $f_0$
despite that the mean value of the dichotomic force equals zero, $\av{f(t)}_T=0$.
Thereby the enhancement of $\av{\dot{x}}$ grows with $x_s$. We emphasize that enhancement of the particle transport is
accompanied by a resonance-like behavior of the effective P{\'e}clet number $\av{\av{\dot{x}}}_T / \av{D_\mathrm{eff}}_T$,
see \figref{fig:Fig4}(b). We find that the latter attains a local maximum at the critical force magnitude $f_0=f_\mathrm{cr}$,
whose value is given by $\av{\dot{x}(-f_0,\Delta\mathcal{P})} / D_\mathrm{eff}(-f_0,\Delta\mathcal{P})$.
While the effective P{\'e}clet number grows with $f_0$ for a symmetric compartment, $x_s=0.5$, it goes to zero for an
asymmetric one, see $x_s=0.99$, as caused by the strong enhancement of the effective diffusivity in triangularly shaped channels \cite{Dagdug2011}.
For $f_0\to \infty$, we expect that $\av{\av{\dot{x}}}_T / \av{D_\mathrm{eff}}_T$ starts to grow since $D_\mathrm{eff}$ converges to unity for infinite strong external bias.

\section{Conclusions}
\label{sec:conclusion}

In summary, we addressed the problem of transport of Brownian particles in a two-dimensional generally asymmetric channel with
periodic confining walls of triangular, zigzag profile. To study rectification mechanisms, we investigated the interplay between
a constant forcing acting on the particles along the channel axis and the Stokes' drag stemming from a pressure-driven flow field.
Along with pure {\it entropic rectification} observed for channels with a broken spatial reflection symmetry in the presence of a constant external bias,
the hydrodynamic flow field paves the  way to efficiently rectify the particle transport  utilizing the effect of {\it hydrodynamically enforced entropic trapping},
being accompanied by a drastically reduced diffusivity. While the usual entropic rectification phenomenon was proposed as the basic
mechanism for efficient separation of particles by generating opposite fluxes for particles of different sizes \cite{Reguera2012},
hydrodynamic flows can be used to achieve lower diffusivity of particles resulting in an increased narrowing of 
distributions for the target particles. The latter feature may prove advantageous  in tailoring more efficient separation
devices as compared to  currently used methods in separating  micro- and/or  nanosized particles.

\section*{Acknowledgments}

\noindent This work has been supported by the Volkswagen Foundation via projects I/83902 (Universit\"at Augsburg) and
I/83903 (Humboldt Universit\"at zu Berlin), the German  cluster of excellence, ``Nanosystems Initiative Munich II''
(NIM II), and the Deutsche Forschungsgemeinschaft via IRTG 1740.

\end{document}